\documentclass[apj]{emulateapj}
\setkeys{Gin}{width=0.48\textwidth}
\usepackage{graphicx}
\newcommand{\hmpc}{\ifmmode{h^{-1}\,\hbox{Mpc}}\else{$h^{-1}$\thinspace Mpc}\fi}

\newcommand{\kms}{\ifmmode{\,\hbox{km\,s}^{-1}}\else {\rm\,km\,s$^{-1}$}\fi}
\newcommand{\msun}{{\rm\,M_\odot}}

\newcommand{\qden}{\msun\, {\rm pc}^{-3} {\rm km}^{-3}{\rm s}^{3}}
\begin{document}
\title{ Early Type Galaxy Core Phase Densities} 
\shorttitle{Early Type Galaxy Core Phase Densities}
\shortauthors{Carlberg \& Hartwick}
\author{R. G. Carlberg}
\affil{Department of Astronomy and Astrophysics, University of Toronto, Toronto, ON M5S~3H4, Canada} \email{carlberg@astro.utoronto.ca }
\author{F. D. A. Hartwick}
\affil{Department of Physics and Astronomy, University of Victoria, Victoria, BC V8W~3P6, Canada}
\email{hartwick@uvic.ca}

\begin{abstract}
Early type galaxies have projected central density brightness profile logarithmic slopes,
$\gamma^\prime$, ranging  from about 0 to 1. 
We show that $\gamma^\prime$ is strongly correlated, $r= 0.83$, with the
coarse grain phase density of the galaxy core, $Q_0\equiv \rho/\sigma^3$.
The $\gamma^\prime$-luminosity correlation is much weaker, $r= -0.51$.
$Q_0$ also  serves to separate the
distribution of power-law profiles, $\gamma^\prime>0.5$ from nearly flat profiles, $\gamma^\prime<0.3$,
although there are many galaxies of intermediate slope, at intermediate $Q_0$, in a volume limited sample.
The transition phase density separating the two profile types is approximately 
$0.003 \qden$, which is also where the relation between $Q_0$ and core mass shows a change in slope, the
rotation rate of the central part of the galaxy increases, and  the ratio of the black hole to core mass increases.
These relations are considered relative to 
the globular cluster inspiral core buildup and binary black hole core scouring mechanisms for 
core creation and evolution. Globular cluster inspiral models have
quantitative predictions that the data support, but no single model yet completely
explains the correlations.
\end{abstract}

\section{INTRODUCTION}

Early type galaxies, ellipticals and S0s, form an impressively regular sequence with luminosity, as seen in the
 Faber-Jackson relation \citep{FJ:76} and generalized in the fundamental plane relations \citep{DD:87,BM:98}. 
The core radius and brightness are basic observational properties which hold clues to the origin of the early type galaxies.
Galaxy merging, star-formation, the presence of
black holes and, in sufficiently dense cores, dynamical friction  
and two-body relaxation, all play some role in creating the cores. 
One widely  considered possibility for the formation and evolution of early type galaxy cores 
is that they are largely a result of stellar dynamical processes in the core
 with gas and star formation playing minor roles.

The coarse grain phase space density can be defined as
$
Q_0 \equiv \rho_0  \sigma_0^{-3}.
$
$Q_0$ is a key dynamical quantity to assess the relative roles of various stellar dynamical processes. 
The local phase density controls the rate of two-body relaxation and dynamical friction, along with the 
 masses of the orbiting bodies \citep{BT:08}. 
Liouville's theorem \citep{Goldstein:02} requires that the fine grain phase density, $F$, is a constant of the motion in a conservative Hamiltonian system, with the coarse grain phase density required to be $Q\le F$ 
(with appropriate velocity space normalization).

The observational description of the central surface brightness distribution of early type galaxies 
began with ground-based telescope data and  used the physically well motivated King model \citep{King:62, King:66, King:78}.  CCDs observations showed that some early type galaxies, particularly those with lower luminosities, had 
central brightness distributions that rose above the constant brightness King model core \citep{Kormendy:85,Lauer:85}
although earlier photographic data had noted this as indicating the presence of a stellar nucleus \citep{Binggeli:84}.
The angular resolution of the Hubble Space Telescope 
showed that what became known as power-law cores were increasingly common with decreasing galaxy luminosity.
\citep{Crane:93, Ferrarese:94, Kormendy:94}. 

The central density profiles  of early type galaxies are classified as being a power-law,  if the negative logarithmic slope 
is greater than 0.5, or, cored, if the  slope is shallower than 0.3
 \citep{Gebhardt:96,Ravindranath:01,Rest:01,Lauer:07b}. 
However, the core parameters and the degree of bimodality of the central slope distribution
depend on the surface brightness fitting  model and the sample definition.

The purpose of this paper is to calculate the phase density of spheroidal systems, focusing on early type galaxies, to consider 
its correlations with core properties.
In particular we examine
to what degree phase density relates to the suggestions that early type cores have a bimodal distribution in their
brightness 
profile slopes. These relations are considered as tests 
of stellar dynamical models for the formation and evolution of early type galaxy cores.

\begin{figure}
\begin{center}
\includegraphics[scale=0.8]{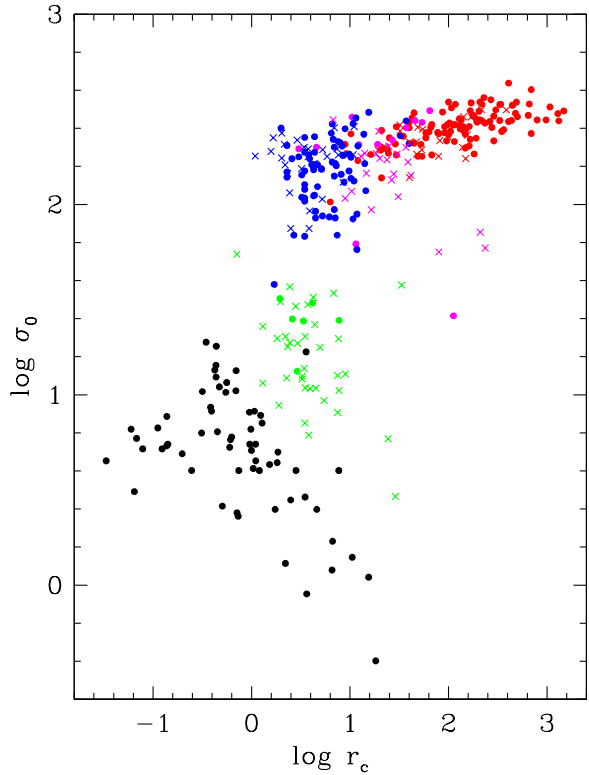}
\end{center}
\caption{The core velocity dispersions and $r_c$ values fitted for early type galaxies from
the \citet{Lauer:07b} (points) and ATLAS3D (crosses) samples.
with cores (red),  intermediate slopes (magenta) and power-law cores (blue),  late type galaxy nuclear star clusters (green points and crosses) and globular clusters (black points). }
\label{fig_sigr}
\end{figure}

\section{The Observational Data}

There are two leading functional forms to fit  a general core brightness profile.
The Nuker formula \citep{Lauer:95,Carollo:97,Lauer:07b}  is designed to describe the central region 
of a galaxy and does not need data at large radii, which is often convenient if there is limited observational coverage.
An alternate, whole galaxy, brightness profile fitting approach is the core-S\'ersic function \citep{GG:03,Graham:03,Trujillo:04,Ferrarese:06,Turner:12, Dullo:12}.
The core-S\'ersic function provides good fits to the entire 
brightness profile of a galaxy at the relatively small cost of a single extra parameter. 
There are significant differences of detail between the two approaches and  no clear consensus 
on best approach has emerged. 

The slope of the brightness profile at the smallest practical radius of observation, $0.1\arcsec$, 
defines the parameter $\gamma^\prime$ of the Nuker approach. The angular definition
introduces a direct distance dependence in the measurement. The Nuker $\gamma^\prime$ is 
general agreement with the core-S\'ersic measurement of the comparable quantity, although with some scatter and
a systematic offset \citep{Dullo:12}.
The fitted break radius of the Nuker profile has limitations in describing the core radius \citep{Carollo:97} 
which led to the development of a transformed quantity, $r_\gamma$, the radius at which 
the Nuker fit has a slope of 0.5,
as the best measure of the core radius. 
\citet{Dullo:12} show that the core-S\'ersic break radius, $r_{b,cS}$, 
is in good agreement with the Nuker $r_\gamma$.

Defining a representative sample of early type galaxies is important when comparing trends within a population.  
The \citet{Lauer:07b} sample is the largest available compilation and is approximately
magnitude limited. A very wide ranging set of
kinematic and dynamical data are available for  the ATLAS3D \citet{Cappellari:11}
sample, which is designed to be a complete volume limited sample. 
Both samples report  Nuker profile fit parameters and
serve as two large, comparable and complementary samples for our analysis.

\subsection{Phase Density Calculation}

To calculate the core coarse grain phase space density requires measurements of 
the core velocity dispersion and core radius, from which other quantities can be derived.  
The different types of systems have a range of density profiles and somewhat different approaches to measurement 
are used, meaning that the phase densities will have small systematic differences between them although we expect
these to much smaller than the large range of $Q_0$ values present.
To obtain a physical mass density in the core we use the King radius relation \citep{BT:08} of an
non-singular isothermal sphere for all systems,
\begin{equation}
\rho_0={9\over{4 \pi G}} {\sigma_0^2\over r_c^2}.
\label{eq_rhoc}
\end{equation}
We recognize that few of the systems are particularly well described as an isotropic isothermal sphere, however 
Equation~\ref{eq_rhoc} provides a uniform basis to calculate basis that will be correct within a factor of a few.
To employ Equation~\ref{eq_rhoc} we need to identify a central velocity dispersion and a measure of the core radius for the systems that we consider. 
Although our primary interest
is the early type galaxies, in which core radii are measured in the same way,
we will make comparisons to phase densities of nuclear star clusters of late type galaxies and globular clusters
in which the measures of core radius are comparable but not identical, 
which will lead to systematic differences, but these will not affect our limited use.

Given our assumptions, 
\begin{equation}
Q_0={166.6\over{\sigma_0 r_c^2}}
\label{eq_Q}
\end{equation} 
for $\sigma_0$ in 
units of $\kms$ and $r_c$ in units of pc. 
$Q_0$ is the approximate phase space density at the core radius 
and normally increases inward from that location.

\subsection{Early type Galaxies}

The \citet{Lauer:07a,Lauer:07b} sample is a large compilation of available HST data, but as an approximately apparent magnitude limited sample contains more luminous galaxies than a volume limited sample
\citep{Lauer:07b,Cote:07}.
In total there are 189 galaxies in the sample that have core profiles and central velocity dispersions.

The ATLAS3D sample \citep{Krajnovic:13} is constructed
to be essentially a complete sample of galaxies more massive than about $6\times 10^9\msun$ within
42 Mpc. 
\citet{Krajnovic:13} provides Nuker fits for the surface brightness profiles.
We remove galaxies with upper limits for $r_\gamma$ from the sample, which reduces the numbers to 74.
We equate $r_\gamma$ with $r_c$ below.
For the ATLAS3D sample we use the values of velocity dispersion at $R_e/8$ as tabulated in \citet{Cappellari:13XX}.

For all early type  galaxies we calculate the core mass from the projected quantities, 
using the relationship $\Sigma_0=2\rho_0 r_c$,
\begin{equation}
M_c=4 \pi \rho_0 r_c\int_{0}^{r_{c}}r(r/r_c)^{-\gamma^{\prime}}\,dr.
\label{eq_mc2d}
\end{equation}
This integrates to $M_c = 4\pi \rho_0 r_c^3/(2-\gamma^\prime)$.

Density profiles are classified as being cored if $\gamma^\prime<0.3$ (red in plots) and power-law if $\gamma^\prime>0.5$ (blue in plots),
with an intermediate type between (magenta in plots) 
\citep{Lauer:07b}. 

\subsection{Globular clusters}

Globular clusters may play a significant role in the formation of galactic cores \citep{TOS:75}.
Accordingly it is interesting to understand where their phase densities fits into the overall sequence \citep{Walcher:05}.
In this work we will assume that the Milky Way globular clusters are 
representative of the group as a whole. For these clusters, core radii and 
central velocity dispersions come from the 2010 edition of the compilation of \citet{Harris:96}. 
In the following figures globular cluster data are plotted as solid black dots.

We use Equation~\ref{eq_rhoc} to calculate the core density. 
We approximate the core to be constant density,  so the core mass is,
\begin{equation}
M_{c}={4 \pi\over 3 } {\rho_0 r_c^3}.
\label{eq_mc}
\end{equation}

\subsection{Disk Galaxy Nuclear Star Clusters}

The data for the nuclear star clusters in disk galaxies comes from the work of
\citet{Boker:04} and \citet{Walcher:05}, who also estimated phase space densities at the half-mass radius. 
There are six galaxies in Table 3 
of \citet{Walcher:05} for which values of the profile fitted $r_{e}$ are available in the 
\citet{Boker:04} paper. Densities for these galaxies were calculated using Equation~\ref{eq_rhoc} 
above but with $r_{c}\equiv r_{e}/0.75$ (that is, the $r_{h}$ of \citet{Walcher:05}).
This identification of core radius effectively considers the entire NSC to be a core. Depending on the
concentration of the core, the phase density can be several times higher.
Under this assumption,
$Q_0$ for each galaxy can then be calculated from the tabulated values of 
$\sigma$. In the 
figures these six galaxies are plotted as solid green dots. For the remaining 
galaxies in the \citet{Boker:04} Table 1 with $r_{e}$ values, core masses 
were calculated from the given luminosities and an assumed $M/L$ of 0.5 which is the 
median value in Table 3 of the \citet{Walcher:05} paper. Velocity dispersions 
were obtained using the following empirically determined relation by fitting to
the data in Table 3 of \citet{Walcher:05},
\begin{equation}
\sigma_0^{2}=0.45{GM_c\over r_e}.
\label{eq_sigc}
\end{equation}
The resulting empirically determined values of $Q_0$ are plotted as green 
crosses in the figures.

The data are plotted in the $r_c$ vs $\sigma_0$ observational plane in Figure~\ref{fig_sigr}. 
Power-law cores are plotted as blue dots and cored galaxies as red dots.
We note that in this plot the
various types of objects have substantial overlap when projected onto any one axis. There is also substantial mass overlap between the various object types.

\begin{figure}
\begin{center}
\includegraphics[scale=0.8]{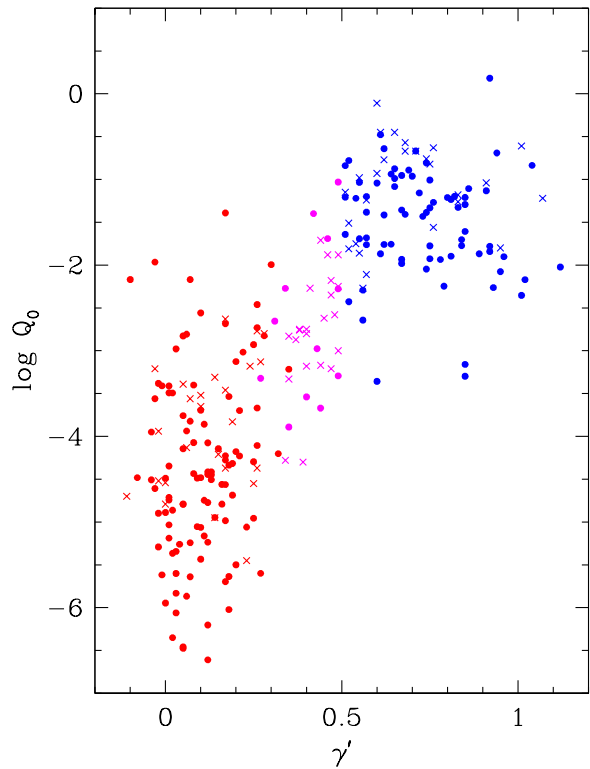}
\end{center}
\caption{The central phase density, $\rho_0/{\sigma_0}^3$, in units of 
$\msun {\rm pc}^{-3} {\rm km}^{-3} {\rm s}^3$ of early type galaxies
as a function of the logarithmic slope of the surface brightness near the center.
The points have the same colors as in Fig.~\ref{fig_sigr}. The \citet{Lauer:07b} sample
is shown with colored dots and the ATLAS3D sample is shown with crosses.
}
\label{fig_qgam}
\end{figure}

\begin{figure}
\begin{center}
\includegraphics[scale=0.8]{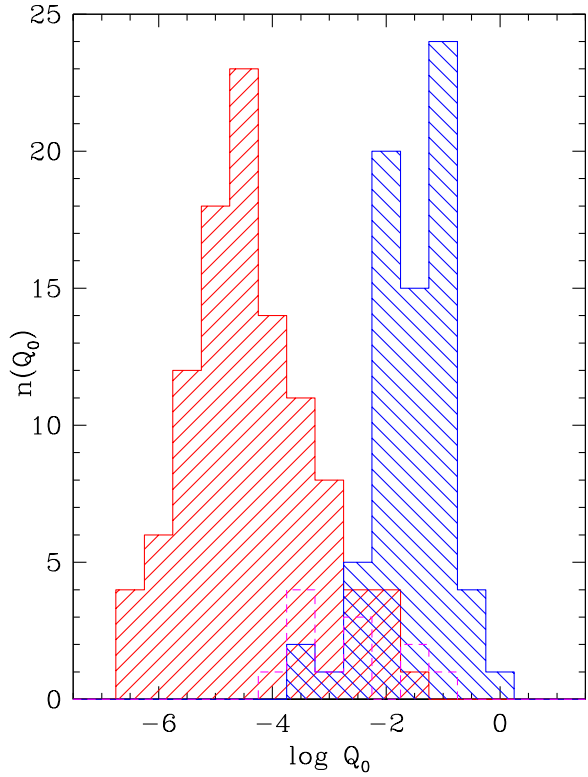}
\end{center}
\caption{The phase space density distribution of early type galaxies classified as
cored (red), intermediate (magenta) and power-law (blue) from the \citet{Lauer:07b} sample. }
\label{fig_hisq}
\end{figure}

\begin{figure}
\begin{center}
\includegraphics[scale=0.8]{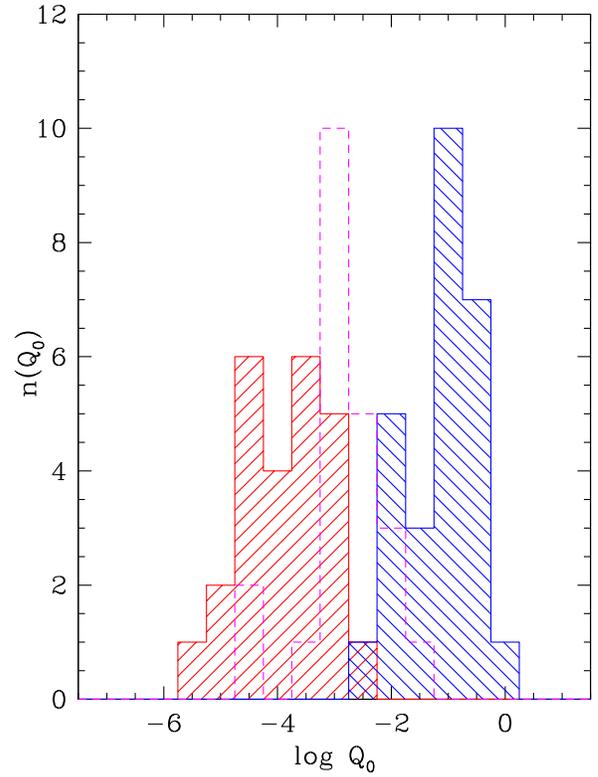}
\end{center}
\caption{Same as Figure~\ref{fig_hisq} but for the volume limited \citet{Krajnovic:13} sample. }
\label{fig_hisqk}
\end{figure}

\begin{figure}
\begin{center}
\includegraphics[scale=0.8]{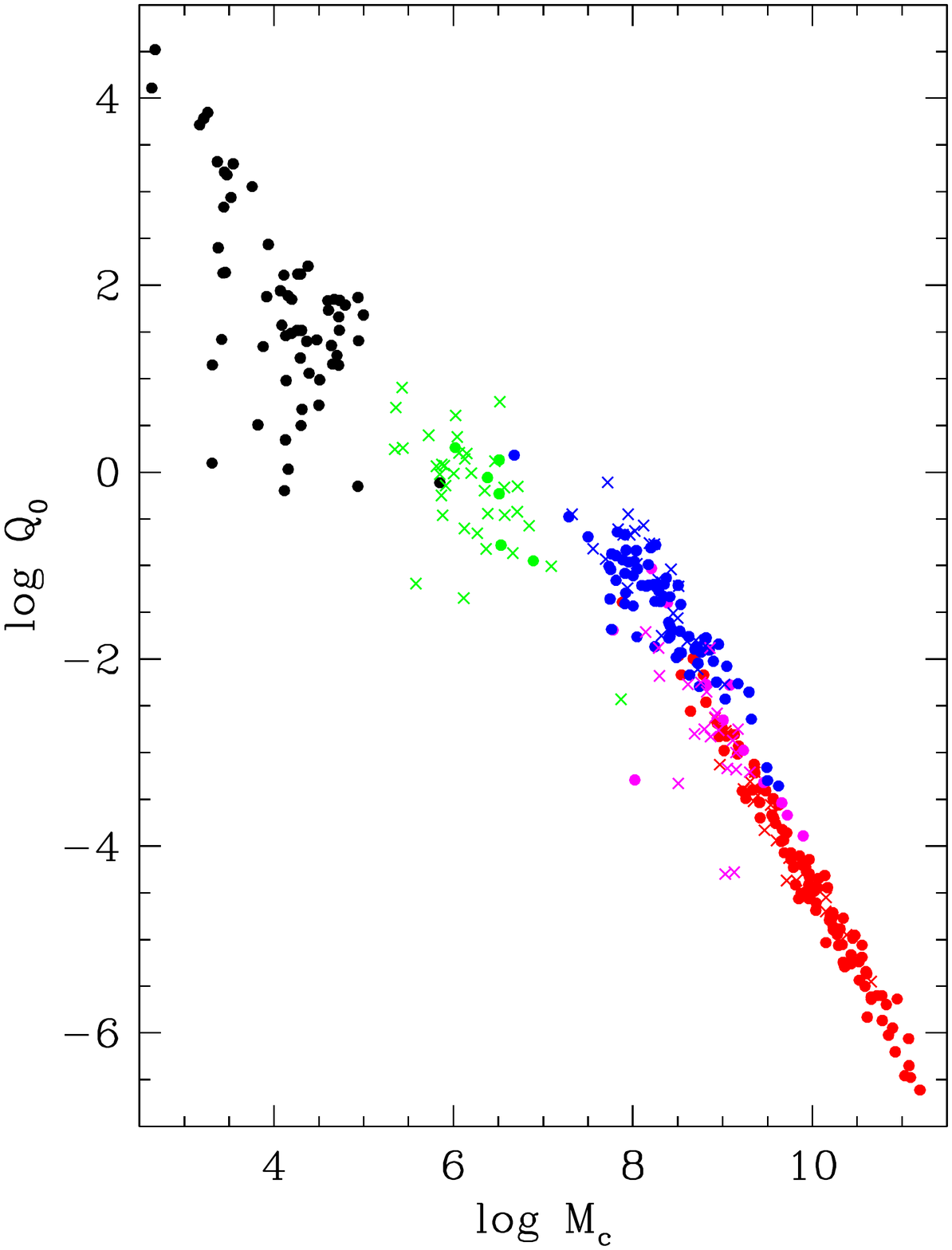}
\end{center}
\caption{The central phase density, $\rho_0/{\sigma_0}^3$, in units of 
$\msun {\rm pc}^{-3} {\rm km}^{-3} {\rm s}^3$. 
The points have the same colors as in Fig.~\ref{fig_sigr}. }
\label{fig_qmc}
\end{figure}

\section{$Q_0$ dependence of Core Slope}

In Figure~\ref{fig_qgam} we plot the core phase densities as a function of their inner brightness profile slope, $\gamma^\prime$,
 for all early
type galaxies for both the \citet{Lauer:07b} and \citet{Krajnovic:13} samples.  
A strong correlation between the log of $Q_0$ and $\gamma^\prime$ is readily visible. 
The Pearson linear correlation coefficients for the $Q_0-\gamma^\prime$ correlation
is 0.78 for the \citet{Lauer:07b} sample and
0.83 for the \citet{Krajnovic:13} sample.
That is,  the core phase space density explains an impressive  69\% ($=r^2$) of the variance in $\gamma^\prime$.
The correlation of $\gamma^\prime$ with luminosity is fairly weak in the ATLAS3D sample,
-0.51, {\it i.~e.} only about 25\% of the variance, where we use the log of the 
$r$ band luminosities, $L_r$, of \citet{Cappellari:13XV}.
The correlation of log $Q_0$ and either log of the stellar mass or $r$ band luminosity
is also comparably weak, $r=$ -0.49 and -0.50, respectively. The weak correlation with the total mass or luminosity of 
the galaxy indicates that although early type cores evolve within their host galaxy, 
the host galaxy does not completely control the resulting core properties.

\begin{figure}
\begin{center}
\includegraphics[scale=0.8]{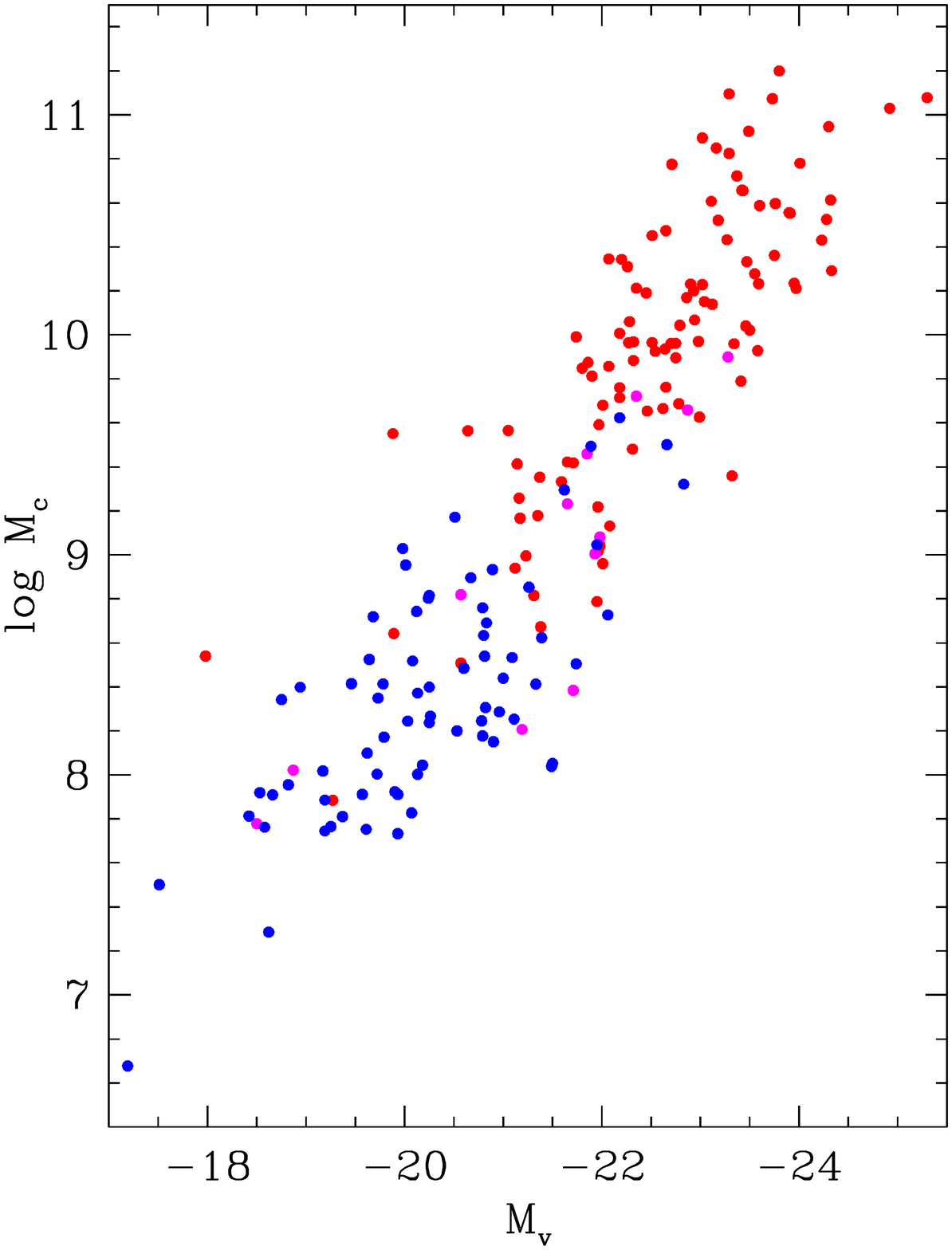}
\end{center}
\caption{The core mass as a function of luminosity for the \citet{Lauer:07b} sample. Symbols are as in Figure~\ref{fig_sigr}.}
\label{fig_mcmv}
\end{figure}

To further explore the correlation of core density profile and the bimodality proposals, 
Figures~\ref{fig_hisq} and \ref{fig_hisqk}
show for the two early type samples the distribution of their core and power-law types, $\gamma^\prime<0.3$ and $\gamma^\prime>0.5$ respectively, as a function of their phase space density.
Remarkably $Q_0$ fairly
 cleanly separates the cored and power-law ellipticals into two almost non-overlapping distributions. 
The relative numbers of core and power-law galaxies are comparable in both samples.
Of the 25 cored and 26 power-law galaxies in the \citet{Krajnovic:13} sample, 
only one from each falls into an overlapping region of $Q_0$ in Figure~\ref{fig_hisqk}.

The intermediate core types, those with $0.3\le \gamma^\prime\le 0.5$, fill in the
$Q_0$ region between the cored and power-law types. The intermediate core slope population 
is relatively large in the \citet{Krajnovic:13} sample, 22 of the 74, with 11 of the 189 in
the \citet{Lauer:07b} sample, likely reflecting the difference between the sample selection criteria.
Although it is true that cored and power-law galaxies do form two 
nearly non-overlapping distributions in $Q_0$,  
a more physical  interpretation is that $\gamma^\prime$ is a continuous variable that is strongly ordered
with the core $Q_0$. 
We note that the transition from power-law to cored galaxies occurs near $Q\simeq 0.003 \qden $.

\begin{figure}
\begin{center}
\includegraphics[scale=0.8]{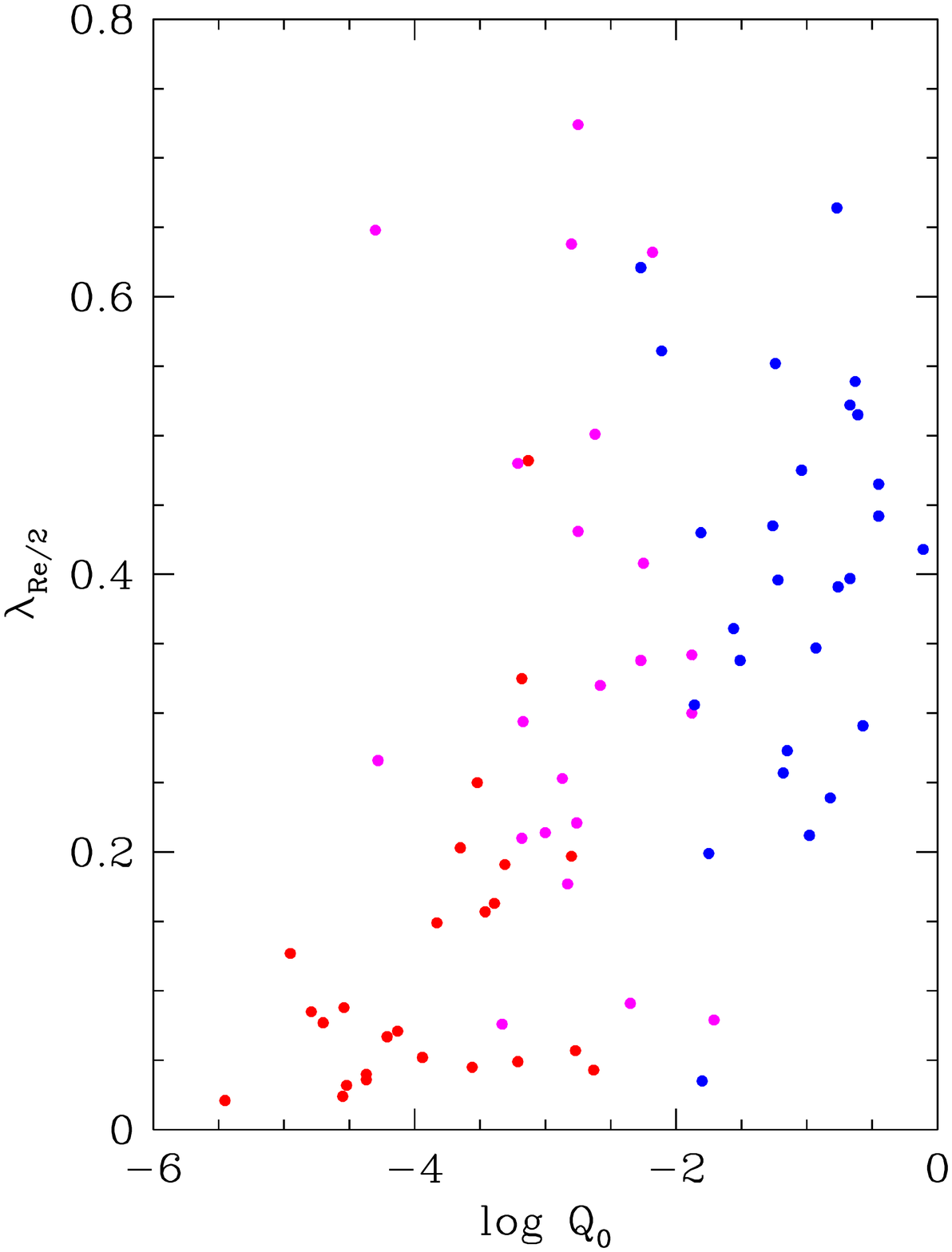}
\end{center}
\caption{The rotation parameter as a function of the phase space density for the ATLAS3D sample. }
\label{fig_lam}
\end{figure}

\section{Dynamical Quantities and $Q_0$}

\subsection{Core Mass}
The core phase densities  of early type galaxies, nuclear star clusters and globular clusters are shown in Figure~\ref{fig_qmc} as a function of the calculated core mass. 
As a group the objects appear to form a single sequence in this particular dynamical space.
There is a distinct break in the slope between cored and power-law galaxies.

We know about what to expect for a  $Q_0-M_c$ relation.
Equations~\ref{eq_Q} and \ref{eq_mc}
give $Q_0\propto \sigma_0^3 M_c^{-2}$. 
If we take a Faber-Jackson relation for the core to be,
$\sigma_0 \propto M_c^\beta$
and use Equation~\ref{eq_rhoc} to eliminate the density, then
\begin{equation}
Q_0\propto M_c^{-2+3\beta}.
\label{eq_qm}
\end{equation}

The power-law cores in Figure~\ref{fig_qmc}
 have a $Q_0-M_c$ slope from 72  points $-1.109 \pm 0.064$ whereas the 105 cored
early type galaxies have $-1.641 \pm 0.025$. 
\citet{KB:13} find Faber-Jackson relationships,  $L_v\propto \sigma_0^n$, where their $n=1/\beta$, 
or $n\simeq 3.74$  for power law galaxies and $8.33$ for cores,
assuming that the core mass is directly proportional to the total luminosity, which is approximately true 
as shown in Figure~\ref{fig_mcmv}.
 Accordingly we would expect $Q_0-M_c$ slopes of -1.20 and -1.64, 
consistent with the fitted relations, for power-law and cored early types, respectively.
The correlation between $Q_0$ and $M_c$ has $r=$ 0.95. However $Q_0$ 
is  better predictor of 
$\gamma^\prime$, with 81\% confidence that the increased correlation of $\gamma^\prime-\log{Q_0}$ over $\gamma^\prime-\log{M_c}$ is significant.

\subsection{Rotation}

\citet{Lauer:12} showed that core type correlates well with the rotation parameter calculated from the sub-sample of ellipticals with rotation maps \citep{Emsellem:04}. 
The kinematic data has been obtained with ground based telescopes so does not have the angular resolution
of the imaging data.
However, it remains of considerable interest to compare the core phase space density with 
rotation properties at a larger radius. 
 Figure~\ref{fig_lam} shows the correlation of the rotation parameter values of \citet{Emsellem:11} 
measured within $R_e/2$ and the ATLAS3D $Q_0$ values. There is a range of rotation at every $Q_0$, with 
a growing maximum rotation with increasing $Q_0$. 
The $Q_0-\lambda_{Re/2}$ correlation coefficient is 0.55. 
Factors other than $Q_0$ dominate the rotation value for an individual galaxy, which is hardly surprising
given that a significant range in the initial angular momentum is expected. However $Q_0$ usefully
indicates the mean rotation, with a value of  $0.38\pm0.03$ above the critical value 
of $Q_0=0.003$ whereas it is $0.21\pm0.03$ below.

We note that the ATLAS3D sample we are using contains 
a number of kinematically distinct cores (KDC) defined 
on the basis that the kinematic axis changes by at least $30\degr$ \citep{Krajnovic:11II} 
For our sub-sample one KDC occurs in an intermediate core type, 
the 8 others are in cored galaxies,  consistent within the small
numbers with the expected fraction of about half the cored galaxies.

\subsection{Dark Matter Fraction}

It is straightforward to undertake a principal component analysis (PCA) using the R project software
(http://www.r-project.org/). 
We remove strongly correlated variables, leaving the PCA with $\gamma^\prime$, the $r$ band luminosity,
and the parameters reported in  \citet{Cappellari:13XX}, the flattening at $Re/2$, the rotation parameter
within $Re/2$, the kinematic/photometric misalignment angle, $\psi$, and $f_{DM}$, the dark matter fraction
inside $Re$.
The PCA finds that the first component provides 49\% of the variance, and three components of seven are 
required to provide 80\% of the variance. 
The PCA confirms that $Q_0$ is the single most strongly correlated variable with $\gamma^\prime$.
However the next most correlated variable is the dark matter fraction, $f_{DM}$, 
as evaluated at $R_e$.
A general linear model fit finds $\gamma^\prime = (0.138\pm 0.015)\log{Q_0} - (0.32\pm0.12)f_{DM} + 
(0.25\pm0.10)\lambda_{Re/2}+0.745$. The $f_{DM}$ term has 
a probability of only 0.8\%  of chance occurrence and the rotation term $\lambda_{Re/2}$ has 
a probability of  1.6\%  of chance occurrence.  However, $\log{Q_0}$ removes about 69\% 
of the variance, whereas both $f_{DM}$ and rotation each account for only about 2.3\%, leaving 26\% unaccounted.
The negative correlation with $\gamma^\prime$ indicates that a lower dark matter fraction is associated
with power-law cores. This is suggestive that somewhat more dissipated galaxies, hence lower dark matter fraction,
are more likely to have steep core profiles at the same $Q_0$, but it is a small effect.

\begin{figure}
\begin{center}
\includegraphics[scale=0.8]{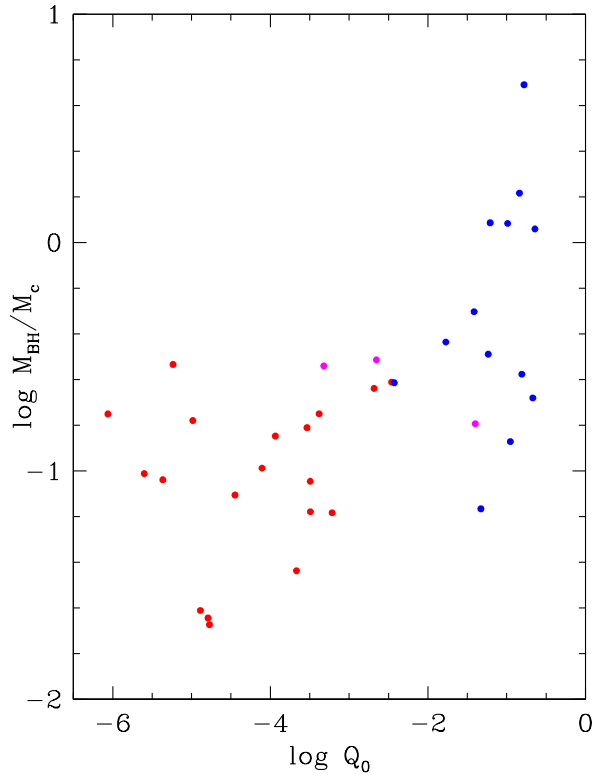}
\end{center}
\caption{The ratio of the central black hole mass to the core mass as a function of $Q$. }
\label{fig_bhr}
\end{figure}

\subsection{Central Super-Massive Black Holes}

Super-massive black holes are closely connected to galaxy cores although the
evolutionary dependence is not entirely clear.
There are well known correlations between the velocity dispersion of ellipticals and their black holes
\citep{Magorrian:98,FM:00,Gebhardt:00}. 
Black hole masses are available in \citet{GrahamScott:13} and \citet{MM:13}. 
Although the galaxies with black-hole masses are a subset of those with core density measurements,
 there is no clear bias other than the size and distance considerations needed to allow a black hole mass measurement.  
Figure~\ref{fig_bhr} shows the ratio of the derived black hole mass to our estimated core masses 
as a function of their core phase densities. 
The high $Q_0$, 
power-law core, galaxies have core masses comparable to the black hole mass,
with a median $M_{BH}/M_c\simeq 0.5$. 
That is, the cores of most galaxies with 
$Q_0 > 0.003\qden$ are within 
the black holes sphere of gravitational influence \citep{MM:01}. 
Cored galaxies have much more massive cores on the average, with a
median  value $M_{BH}/M_c \simeq 0.1$. 


\section{Implications of Core Phase Densities}

\subsection{Dynamical Times in Dense Cores}

The phase space density determines the time for two-body relaxation \citep{BT:08}.  
We adopt the parameter choices of \citet{Merritt:13}, his Equation 5.61, to find,
\begin{equation}
t_{2}  = {1.1 \times 10^9\over Q}  {\msun\over m} {15\over{\ln{\Lambda}}} {\rm yr}.
\label{eq_tr}
\end{equation}
A relaxation  time of $10^9$ yr for a population of $m=1 \msun$ stars occurs for $	Q=1.1$. 
The power-law cores containing black holes shown in 
 Figure~\ref{fig_bhr}, have median $Q_0\simeq 0.1$, 
in which case interaction with two-body relaxation of stars around the black-hole
likely plays a significant role in building and maintaining a power-law core 
\citep{MerrittSzell:06}.

The time scale for dynamical friction depends linearly on the phase space density and the mass of the inspiraling satellite  \citep{BT:08}.
We again use the parameters of \citet{Merritt:13}, his Equation 5.32,  to give the
radial dependence of the dynamical friction timescale,
\begin{equation}
t_{df}(r)={1.4 \times 10^{10}\over {Q(r) }} {0.35\over {G(x)}} {\msun\over m} {10\over{\ln{\Lambda}}} {\rm yr}
\label{eq_tdf}
\end{equation}
where $G(v/\sigma)$ is a slowly varying function with a value of 0.347 for $v/\sigma=\sqrt{2}$, roughly as expected for an inspiralling satellite.
For a  somewhat heavy $3\times 10^5\msun$ globular cluster  to spiral in within a $10^{10}$ years requires 
that it orbit in a region with a local $Q >4\times 10^{-6}\qden$. 
Therefore, the most massive early type galaxies, those with cores 
more massive than about $10^{10}\msun$ (Figure~\ref{fig_qmc})
only allow globulars more massive than $3\times 10^5\msun$ 
to spiral into the center.

\subsection{Core Phase Density Evolution Pathways}

A successful core formation and evolution scenario needs to offer a way to understand the correlations
presented here
between the core phase density and its brightness profile, rotation and black hole mass.
High redshift progenitor galaxies may largely form with power-law cores through dissipative
star formation processes \citep{Loose:82}
although current specific models appear to create too steep a Faber-Jackson relation \citep{McLaughlin:06,Antonini:13}.
A widely investigated approximation, which we will follow, is
to assume that the stars in the cores of early type galaxies were largely created elsewhere 
and brought to the core through merging and dynamical friction and possibly had their
density profile altered through two-body relaxation.
Here we briefly examine to what degree
these stellar dynamical processes can account for the phase space correlations of core properties.
Two widely discussed and quite distinct ideas  are either that early type 
galaxies began with large low density cores
which were subsequently built up, or, they were formed with power law cores, which were subsequently 
scoured out in the more massive galaxies. Both mechanisms may be at work.

\begin{figure}
\begin{center}
\includegraphics[scale=0.7]{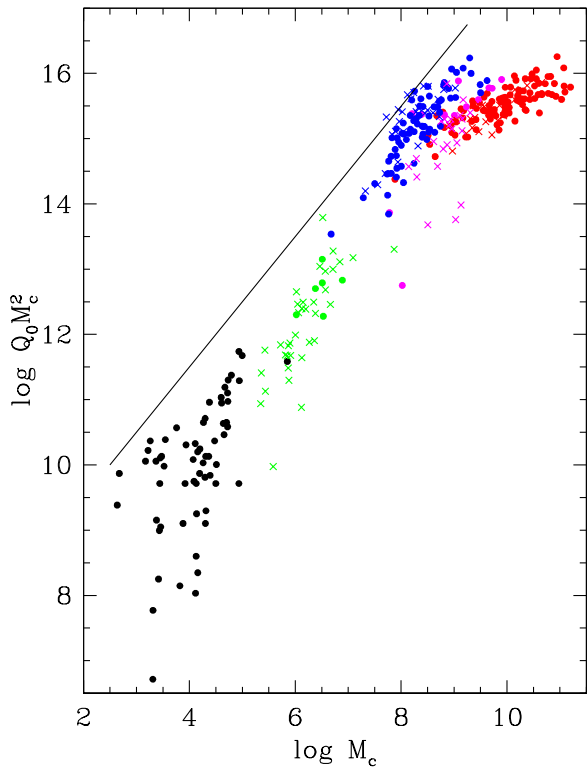}
\end{center}
\caption{The  phase space density scaled with the core mass squared, in which mergers
move as $Q_0M_c^2 \propto M_c^{3\beta}$. 
The symbols are as in Figure~\ref{fig_sigr}.
A line of constant density,  $Q_0 M_c^2\propto \rho^{1/2}M_c$, 
is shown.}
\label{fig_qm2m}
\end{figure}

The slope of the power-law cores in the $Q_0-M_c$ diagram, Figure~\ref{fig_qmc}, appears to be 
a continuation of the trendline from globular clusters through Nuclear Star Clusters. 
Precisely the same data are multiplied with its core mass squared to remove the basic mass dependence
of 
Equation~\ref{eq_qm}. The rescaled results are displayed in Figure~\ref{fig_qm2m} which helps make the idea of two
sequences clearer. Shown on the plot is a line of constant density, under the
self gravitating  virial assumption, which gives
$\sigma^2 \propto M_c^{2/3} \rho^{1/3}$ and therefore $Q_0M_c^2 \propto \rho^{1/2} M_c$.
 
N-body simulations for the bulk of the galaxy find  $\beta \simeq 0.3$ for equal mass mergers 
and $\beta\simeq -0.3$ for minor mergers \citep{Naab:09,Hilz1}, although these values
will not necessarily strictly apply to the cores. Recall that we have 
translated the Faber-Jackson relation to $Q_0M_c^2\propto M_c^{3 \beta}$.
The shallow trend of the cored early types in Figure~\ref{fig_qm2m} is highly suggestive that 
{\it major} merging dominates the evolution of these cores, with some admixture of minor merging to flatten out the relation,
as has been noted from their Faber-Jackson relations \citep{KB:13}
If the merging were predominantly minor mergers it would cause  $Q_0M_c^2$ to decline with increasing core mass,
which is not seen. 
It is also suggestive that  a steeper trend continues through power-law cores 
as has been previously noted \citep{Walcher:05, Cote:07,Glass:11}. The globular cluster core densities are set through star formation processes and subsequent dynamical
evolution. 
The globular cluster inspiral process will dissolve the clusters when the mean interior density of galaxy core and the 
cluster are equal, that is, the mass buildup will tend
to push the increasing mass objects along a roughly constant density line, 
\citep{CapuzzoDolcetta:08,Hartmann:11}. A sequence of infall events leads to a modest increase
in central density \citep{Antonini:12}. The presence of a central black hole
complicates the process but the general trends remain \citep{Antonini:13}.

\subsubsection{The Power-law Core Buildup Scenario}

Although not necessarily the only mechanism which creates shallow cores, one
interesting possibility is that most of the stars in early type galaxies may have been formed in galactic disks,
either isolated or in star bursts associated with merging. Subsequent merging of the 
stellar components leads to a core phase density that is no higher than the 
phase density of the central regions of the merging disks \citep{TT:72,Barnes:88,Cox:06}.

\citet{Genzel:11, Genzel:14} presents observational measurements of star forming galaxies near redshift two.
The phase space density of a disk is $Q_d= \Sigma/(2h\sigma^3)$, using our normalization and where $\Sigma$ is
the local surface density, $h$ the scale height, and $\sigma$ the velocity dispersion which we will take to be isotropic
for simplicity.
Taking high redshift disk parameters of $\Sigma\simeq2-3\times 10^3 \msun {\rm pc}^{-2}$ (with some fraction  as stars), 
$z_0\simeq 300$ pc, and $\sigma\simeq80 \kms$ \citep{Genzel:11}, 
which gives a representative central disk $Q_0\simeq 3\times 10^{-6}\qden$. 
The Milky Way has a comparable central phase density.
Merging such stellar disks would create an initial  core 
$Q_0\simeq 10^{-6}\qden$, if about 1/3 of the gas eventually 
turns into stars and the rest of the central
gas is driven away.  Dissipationless merging of the stellar components of such disks would create galaxies with core phase densities comparable to the most massive, lowest $Q_0$, early type galaxies  \citep{Carlberg:86}.
For disks in the same potential, but with lower surface mass densities that phase density will be higher.
That is, in a plane parallel sheet  $\pi G \Sigma h = \sigma^2$,
$Q_d \propto \Sigma^2 \sigma^{-5}$. 
To keep the Toomre disk stability parameter \citep{Toomre:64}
 near unity, requires that the velocity dispersion adjust in proportion to the 
surface mass density, $\sigma \propto \Sigma$. Therefore $Q_d \propto \sigma^{-3}$. The minimum velocity dispersion of a galactic disk is approximately 10 \kms, due to 
stirring of molecular clouds and internal motions of dissolving star clusters,
so the highest phase density that merging disks would create would be  $\simeq 10^{-3} \qden$. 
Therefore a pure disk merger scenario could account for the core phase densities of 
all the cored, $\gamma^\prime<0.3$, early type galaxies but 
cannot account for the higher phase densities of the early type-galaxies with power-law cores.
This is not to dismiss the important role that core scouring likely also plays, as discussed below.

Assuming that all early type galaxies do begin with fairly flat cores, 
a widely discussed mechanism for subsequent core buildup is that
globular clusters can be dragged into the centers of galaxies 
\citep{TOS:75}.  Although the current globular cluster population is insufficient to provide the required mass,
it is likely that at high redshift significantly more high mass clusters were present, which
evolved under the action of various dynamical processes that have been studied and tested extensively 
\citep{MW:97a,MW:97b,Fall:01,MF:08,Gieles:09,Larsen:09,Chandar:10,FC:12}.  
Allowing for the evolution of the globular cluster population 
can build up all of the nuclear star cluster mass in galaxies with stellar masses below about 
$10^{11} \msun$ \citep{Antonini:12,Antonini:13}. 

The phase density can be used for a rough calculation of the accreted globular cluster mass. 
As discussed with Equation~\ref{eq_tdf}, 
the phase density at the maximum radius from which globular clusters can spiral in 
is $Q\gtrsim 4\times 10^{-6}\qden$, at which phase density the 
 interior stellar mass of $\simeq 2.0\times 10^{10}\msun$,
which can be read off Figure~\ref{fig_qmc}.
For the greatly enhanced progenitor 
globular cluster to stellar mass fraction of 0.04 that \citet{GOT:13} suggest, the accreted mass will be 
$8 \times 10^8 \msun$ which spirals down the center to build a core of that mass. 
Referring to Figure~\ref{fig_qmc} we see that this mass is approximately where the transition
value of $Q_0\simeq 0.003$ that separates power-law from cored early type galaxies occurs. 
The globular cluster
population enhancement, which only needs to apply to the inner kiloparsec or so, is about
a factor of tend over current epoch galaxies have an average of approximately 0.003 of their stellar mass in globular clusters \citep{Harris:13}.
Outside the core the stellar density distribution that is providing the friction generally falls faster than the core mass - core radius relation so smaller cores, which have higher $Q_0$, will bring in fewer clusters. 
Since the available globular cluster mass is proportional to the galaxy mass, the inspiralled globular cluster mass will scale in proportion to the core mass, Figure~\ref{fig_mcmv}. 
A second constraint on the progenitor globular cluster population is that it must have significant net rotation 
to produce the rotation of power law cores, Figure~\ref{fig_lam}.

The detailed calculations of
\citet{GOT:13} predict that the excess mass in the core relative to the galaxy mass 
should be proportional to the inverse square galaxy mass, comparable to the excess 
mass-velocity dispersion relation of \citet{Antonini:13}.
To test this prediction we calculate $\Delta M_c$ as the difference between Equations~\ref{eq_mc} and a constant 
density core. We define $\Delta M_c$ as $M_c(\gamma^\prime)-M_c(\gamma^\prime=0)$. 
The  correlation between the logarithms of $\Delta M_c/M_\star$ and $M_\star$, the stellar mass of the galaxy, 
is a very weak $r= -0.33$ for all the ATLAS3D galaxies.   Restricting the fit to the galaxies with
$\gamma^\prime>0.3$, {\it i. e.} all non-cored galaxies, the slope of the $\log{\Delta M_c/M_\star}-\log{M_\star}$ relation 
is $-0.52\pm0.13$, in accord
with the predicted inverse square root relation \citep{GOT:13}
but the correlation 
 rises to only a marginal -0.49. 
Although it seems inevitable that a substantial number of globular clusters have 
spiralled into galactic cores, the predicted relation between $\Delta M_c$ and $M_\star$ accounts for only  about 11\% of the variance. Of course galaxy to galaxy differences in formation history, central globular cluster numbers, and the
specific properties of the globulars from one galaxy to another could account for the scatter.

\subsubsection{The Core Scouring Scenario}

All early type galaxies could be created with fairly steep power-law cores, which then requires
a mechanism to hollow out the cores in larger galaxies.
Super-massive black holes pairs that come together after two progenitor galaxies merge
will quickly sink into the core  \citep{Begelman:80}. 
Stars that encounter the pair can be ejected from the core to``scour'' 
out of a power law core and produce a much shallower central density profile \citep{Ebisuzaki:91,Faber:97,MM:01}.
The process preferentially depletes stars on radial orbits to leave a relatively 
tangential velocity ellipsoid in the core  \citep{Quinlan:97,MM:01,Antonini:12}.
Kinematic observations of a sample of cored early type galaxies do show the expected
signature \citep{Thomas:13}.
Simulations have shown that a single binary black hole pair ejects only a few times the 
resulting merged black hole mass, so the  small $M_{BH}/M_c$ ratios of cored galaxies require multiple mergers
to successfully scour the core \citep{Faber:97,MM:01,Merritt:06,GualandrisMerritt:08,Dullo:13}.

\citet{KB:13} present observational arguments for a scenario  in which subsequent nearly dissipationless
major merging dominates the formation of the most massive early type galaxies, as indicated by
the very slow rise of central velocity dispersion with increasing mass.
Their Faber-Jackson relations for core and power-law galaxies can be translated into the core phase density - core mass
 relation which we present. The core scouring model at present does not address the strong correlation between 
the core profile slope $\gamma^\prime$ and the core phase density.  Since power-law core galaxies also contain 
black holes they too should have been subject to core-scouring, 
but many of them have sufficiently high core phase densities that a power law core 
can be maintained or rebuilt \citep{BW:76,MerrittSzell:06,Merritt:09}.
Core scouring certainly plays some role which can vary from galaxy between creating the core to largely 
one of maintaining a shallow profile.

\section{CONCLUSIONS}

The coarse grain core phase density, $Q_0$, of early type galaxies provides dynamical insight into 
the properties of the galactic cores and serves as a dynamical ordering parameter.
The core phase density and the slope of the core brightness profile, 
$\gamma^\prime$, are very strongly correlated, $r= 0.83$. 
We also find that for the standard definitions of cored and power-law
profiles, $\gamma^\prime < 0.3$ and $\gamma^\prime >0.5$, respectively, then 
$Q_0$ does a very good job of separating the two distributions, although there is a substantial 
intermediate core slope population between the two.
The  transition from power-law and cored elliptical occurs around $Q_0\simeq  0.003 \qden$. 
$Q_0$ correlates significantly but less strongly with the galaxy rotation, 
the ratio of the central black hole mass to the core mass, and 
quite weakly with the fraction of dark matter inside the effective radius.

Primordial cores could have been built as power-laws through dissipative star formation or as relatively low density cored profiles through disk merging. 
The two main models for subsequent core evolution, globular cluster inspiral and core scouring by black hole binary pairs,
are considered relative to the phase space density correlations. We find that the predicted
relation between excess mass above a flat core and galaxy stellar mass is present in the power-law and
intermediate slope cores but with considerable scatter.
In a diagram of $Q_0$ weighted with the $M_c^2$ as a function of $M_c$ the power-law cores are the
high mass end  of
a sequence from globular clusters through Nuclear Star Clusters. The sequence has slowly rising
core density with mass, roughly as 
simulations of tidal dissolution of globular clusters have found.
The cored early types are on a much shallower relation,
about as would be expected from major mergers.
The transition $Q_0$ between power-law and cored early types can be understood as the maximum possible power-law 
core that can be built from the greatly enhanced globular cluster population at high redshift 
present in the largest cored early type galaxy.
Core scouring clearly plays a role in maintaining fairly flat core profiles, 
but offers no ready explanation for the fairly abrupt disappearance of cored galaxies at a phase density of $Q_0=0.003\qden$.

We conclude that no single model completely explains the core slope-phase density relationship, 
although a combination of stellar dynamical effects seems likely to be the mechanism 
which leaves an indicative signature in the 
strong correlation with core phase density.

\acknowledgements

This research was supported by CIFAR and NSERC Canada. We thank an anonymous referee and 
Davor Krajnovi{\'c} for comments.


\begin{thebibliography}{99}


\bibitem[Antonini et al.(2012)]{Antonini:12} Antonini, F., Capuzzo-Dolcetta, R., Mastrobuono-Battisti, A., \& Merritt, D.\ 2012, \apj, 750, 111 

\bibitem[Antonini(2013)]{Antonini:13} Antonini, F.\ 2013, \apj, 763, 62 

\bibitem[Barnes(1988)]{Barnes:88} Barnes, J.~E.\ 1988, \apj, 331, 
699 

\bibitem[Bahcall \& Wolf(1976)]{BW:76} Bahcall, J.~N., \& Wolf, R.~A.\ 1976, \apj, 209, 214 

\bibitem[Begelman et al.(1980)]{Begelman:80} Begelman, M.~C., Blandford, R.~D., \& Rees, M.~J.\ 1980, \nat, 287, 307 



\bibitem[Binggeli et al.(1984)]{Binggeli:84} Binggeli, B., Sandage, A., \& Tarenghi, M.\ 1984, \aj, 89, 64 

\bibitem[Binney \& Merrifield(1998)]{BM:98} Binney, J., \& Merrifield, M.\ 1998, Galactic astronomy,  Princeton University Press, Princeton

\bibitem[Binney \& Tremaine(2008)]{BT:08} Binney, J., \& Tremaine, S.\ 2008, Galactic Dynamics: Second Edition,  Princeton University Press, Princeton

\bibitem[B{\"o}ker et al.(2004)]{Boker:04} B{\"o}ker, T., Sarzi, M., McLaughlin, D.~E., et al.\ 2004, \aj, 127, 105 



\bibitem[Carlberg(1986)]{Carlberg:86} Carlberg, R.~G.\ 1986, \apj, 310, 593 

\bibitem[Carollo et al.(1997)]{Carollo:97} Carollo, C.~M., Franx, M., Illingworth, G.~D., \& Forbes, D.~A.\ 1997, \apj, 481, 710 

\bibitem[Cappellari et al.(2011)]{Cappellari:11} Cappellari, M., Emsellem, E., Krajnovi{\'c}, D., et al.\ 2011, \mnras, 413, 813 

\bibitem[Cappellari et al.(2013)]{Cappellari:13XV} Cappellari, M., Scott, N., Alatalo, K., et al.\ 2013, \mnras, 432, 1709 


\bibitem[Cappellari et al.(2013)]{Cappellari:13XX} Cappellari, M., McDermid, R.~M., Alatalo, K., et al.\ 2013, \mnras, 432, 1862 

\bibitem[Capuzzo-Dolcetta \& Miocchi(2008)]{CapuzzoDolcetta:08} Capuzzo-Dolcetta, R., \& Miocchi, P.\ 2008, \apj, 681, 1136 

\bibitem[Chandar et al.(2010)]{Chandar:10} Chandar, R., Fall, S.~M., \& Whitmore, B.~C.\ 2010, \apj, 711, 1263 


\bibitem[C{\^o}t{\'e} et al.(2006)]{Cote:06} C{\^o}t{\'e}, P., Piatek, S., Ferrarese, L., et al.\ 2006, \apjs, 165, 57 

\bibitem[C{\^o}t{\'e} et al.(2007)]{Cote:07} C{\^o}t{\'e}, P., Ferrarese, L., Jord{\'a}n, A., et al.\ 2007, \apj, 671, 1456 



\bibitem[Cox et al.(2006)]{Cox:06} {Cox, T.~J., Dutta, S.~N., Di Matteo, T., et al.\ 2006, \apj, 650, 791} 

\bibitem[Crane et al.(1993)]{Crane:93} Crane, P., Stiavelli, M., King, I.~R., et al.\ 1993, \aj, 106, 1371 



\bibitem[Djorgovski \& Davis(1987)]{DD:87} Djorgovski, S., \& Davis, M.\ 1987, \apj, 313, 59 



\bibitem[Dullo \& Graham(2012)]{Dullo:12} Dullo, B.~T., \& Graham, A.~W.\ 2012, \apj, 755, 163 
\bibitem[Dullo \& Graham(2013)]{Dullo:13} Dullo, B.~T., \& Graham, A.~W.\ 2013, \apj, 768, 36 

\bibitem[Ebisuzaki et al.(1991)]{Ebisuzaki:91} Ebisuzaki, T., Makino, J., \& Okumura, S.~K.\ 1991, \nat, 354, 212 

\bibitem[Emsellem et al.(2004)]{Emsellem:04} Emsellem, E., Cappellari, M., Peletier, R.~F., et al.\ 2004, \mnras, 352, 721 

\bibitem[Emsellem et al.(2007)]{Emsellem:07} Emsellem, E., Cappellari, M., Krajnovi{\'c}, D., et al.\ 2007, \mnras, 379, 401 
\bibitem[Emsellem et al.(2011)]{Emsellem:11} Emsellem, E., Cappellari, M., Krajnovi{\'c}, D., et al.\ 2011, \mnras, 414, 888 


\bibitem[Faber \& Jackson(1976)]{FJ:76} Faber, S.~M., \& Jackson, R.~E.\ 1976, \apj, 204, 668 

\bibitem[Faber et al.(1997)]{Faber:97} Faber, S.~M., Tremaine, S., Ajhar, E.~A., et al.\ 1997, \aj, 114, 1771 


\bibitem[Fall \& Zhang(2001)]{Fall:01} Fall, S.~M., \& Zhang, Q.\ 2001, \apj, 561, 751 


\bibitem[Fall \& Chandar(2012)]{FC:12} Fall, S.~M., \& Chandar, R.\ 2012, \apj, 752, 96 

\bibitem[Ferrarese et al.(1994)]{Ferrarese:94} Ferrarese, L., van den Bosch, F.~C., Ford, H.~C., Jaffe, W., 
\& O'Connell, R.~W.\ 1994, \aj, 108, 1598 

\bibitem[Ferrarese \& Merritt(2000)]{FM:00} Ferrarese, L., \& Merritt, D.\ 2000, \apjl, 539, L9 

\bibitem[Ferrarese et al.(2006)]{Ferrarese:06} Ferrarese, L., C{\^o}t{\'e}, P., Jord{\'a}n, A., et al.\ 2006, \apjs, 164, 334 



\bibitem[Gebhardt et al.(1996)]{Gebhardt:96} Gebhardt, K., Richstone, D., Ajhar, E.~A., et al.\ 1996, \aj, 112, 105 

\bibitem[Gebhardt et al.(2000)]{Gebhardt:00} Gebhardt, K., Bender, R., Bower, G., et al.\ 2000, \apjl, 539, L13 

\bibitem[Genzel et al.(2011)]{Genzel:11} Genzel, R., Newman, S., Jones, T., et al.\ 2011, \apj, 733, 101 

\bibitem[Genzel et al.(2014)]{Genzel:14} Genzel, R., F{\"o}rster Schreiber, N.~M., Lang, P., et al.\ 2014, \apj, 785, 75 

\bibitem[Glass et al.(2011)]{Glass:11} Glass, L., Ferrarese, L., C{\^o}t{\'e}, P., et al.\ 2011, \apj, 726, 31 

\bibitem[Gieles(2009)]{Gieles:09} Gieles, M.\ 2009, \mnras, 394, 2113 

\bibitem[Gnedin et al.(2013)]{GOT:13} Gnedin, O.~Y., Ostriker, J.~P., \& Tremaine, S.\ 2013, arXiv:1308.0021 

\bibitem[Goldstein et al.(2002)]{Goldstein:02} Goldstein, H., Poole,  C., \& Safko, J.\ 2002, Classical mechanics (3rd ed.)  Addison-Wesley, San Francisco

\bibitem[Graham \& Guzm{\'a}n(2003)]{GG:03} Graham, A.~W., \& Guzm{\'a}n, R.\ 2003, \aj, 125, 2936 


\bibitem[Graham et al.(2003)]{Graham:03} Graham, A.~W., Erwin, P., Trujillo, I., \& Asensio Ramos, A.\ 2003, \aj, 125, 2951 
\bibitem[Graham \& Scott(2013)]{GrahamScott:13} Graham, A.~W., \& Scott, N.\ 2013, \apj, 764, 151 

\bibitem[Gualandris \& Merritt(2008)]{GualandrisMerritt:08} Gualandris, A., \& Merritt, D.\ 2008, \apj, 678, 780 

\bibitem[Harris(1996)]{Harris:96} Harris, W.~E.\ 1996, \aj, 112, 1487 

\bibitem[Harris et al.(2013)]{Harris:13} Harris, W.~E., Harris, G.~L.~H., \& Alessi, M.\ 2013, \apj, 772, 82 

\bibitem[Hartmann et al.(2011)]{Hartmann:11} Hartmann, M., Debattista, V.~P., Seth, A., Cappellari, M., 
\& Quinn, T.~R.\ 2011, \mnras, 418, 2697 


\bibitem[Hilz et al.(2012)]{Hilz1} Hilz, M., Naab, T.,Ostriker, J.~P., et al.\ 2012, \mnras, 425, 3119 

\bibitem[King(1966)]{King:66} King, I.~R.\ 1966, \aj, 71, 64 

\bibitem[King(1962)]{King:62} King, I.\ 1962, \aj, 67, 471 

\bibitem[King(1978)]{King:78} King, I.~R.\ 1978, \apj, 222, 1 

\bibitem[Kormendy(1985)]{Kormendy:85} Kormendy, J.\ 1985, \apjl, 292, L9 


\bibitem[Krajnovi{\'c} et al.(2011)]{Krajnovic:11II} Krajnovi{\'c}, D., Emsellem, E., Cappellari, M., et al.\ 2011, \mnras, 414, 2923 

\bibitem[Kormendy et al.(1994)]{Kormendy:94} Kormendy, J., Dressler, A., Byun, Y.~I., et al.\ 1994, European Southern Observatory Conference and Workshop Proceedings, 49, 147 

\bibitem[Kormendy \& Bender(2013)]{KB:13} Kormendy, J., \& Bender, R.\ 2013, \apjl, 769, L5 

\bibitem[Krajnovi{\'c} et al.(2013)]{Krajnovic:13} Krajnovi{\'c}, D., Karick, A.~M., Davies, R.~L., et al.\ 2013, \mnras, 433, 2812 

\bibitem[Larsen(2009)]{Larsen:09} Larsen, S.~S.\ 2009, \aap, 494, 539 

\bibitem[Lauer(1985)]{Lauer:85} Lauer, T.~R.\ 1985, \apj, 292, 104 



\bibitem[Lauer et al.(1995)]{Lauer:95} Lauer, T.~R., Ajhar, E.~A., Byun, Y.-I., et al.\ 1995, \aj, 110, 2622 

 
\bibitem[Lauer et al.(2007a)]{Lauer:07a} Lauer, T.~R., Gebhardt, K., Faber, S.~M., et al.\ 2007, \apj, 664, 226 

\bibitem[Lauer et al.(2007b)]{Lauer:07b} Lauer, T.~R., Faber, S.~M., Richstone, D., et al.\ 2007, \apj, 662, 808 
\bibitem[Lauer(2012)]{Lauer:12} Lauer, T.~R.\ 2012, \apj, 759, 64
\bibitem[Loose et al.(1982)]{Loose:82} Loose, H.~H., Kruegel, E., \& Tutukov, A.\ 1982, \aap, 105, 342 





\bibitem[Magorrian et al.(1998)]{Magorrian:98} Magorrian, J., Tremaine, S., Richstone, D., et al.\ 1998, \aj, 115, 2285 

\bibitem[McConnell \& Ma(2013)]{MM:13} McConnell, N.~J., \& Ma, C.-P.\ 2013, \apj, 764, 184 

\bibitem[McLaughlin et al.(2006)]{McLaughlin:06} McLaughlin, D.~E., King, A.~R., \& Nayakshin, S.\ 2006, \apjl, 650, L37 

\bibitem[McLaughlin \& Fall(2008)]{MF:08} McLaughlin, D.~E., \& Fall, S.~M.\ 2008, \apj, 679, 1272 


\bibitem[Merritt(2006)]{Merritt:06} Merritt, D.\ 2006, \apj, 648, 976 

\bibitem[Merritt \& Szell(2006)]{MerrittSzell:06} Merritt, D., \& Szell, A.\ 2006, \apj, 648, 890 


\bibitem[Merritt(2009)]{Merritt:09} Merritt, D.\ 2009, \apj, 694, 959 

\bibitem[Merritt(2013)]{Merritt:13} Merritt, D.\ 2013, Dynamics and Evolution of Galactic Nuclei,  Princeton University Press, Princeton


\bibitem[Milosavljevi{\'c} \& Merritt(2001)]{MM:01} Milosavljevi{\'c}, M., \& Merritt, D.\ 2001, \apj, 563, 34 

\bibitem[Murali \& Weinberg(1997a)]{MW:97a} Murali, C., \& Weinberg, M.~D.\ 1997, \mnras, 288, 749 

\bibitem[Murali \& Weinberg(1997b)]{MW:97b} Murali, C., \& Weinberg, M.~D.\ 1997, \mnras, 291, 717 

\bibitem[Naab et al.(2009)]{Naab:09} Naab, T., Johansson, P.~H., \& Ostriker, J.~P.\ 2009, \apjl, 699, L178 


\bibitem[Quinlan \& Hernquist(1997)]{Quinlan:97} Quinlan, G.~D., \& Hernquist, L.\ 1997,  New Astronomy, 2, 533 

\bibitem[Ravindranath et al.(2001)]{Ravindranath:01} Ravindranath, S., Ho, L.~C., Peng, C.~Y., Filippenko, A.~V., 
\& Sargent, W.~L.~W.\ 2001, \aj, 122, 653 


\bibitem[Rest et al.(2001)]{Rest:01} Rest, A., van den Bosch, F.~C., Jaffe, W., et al.\ 2001, \aj, 121, 2431 




\bibitem[Thomas et al.(2013)]{Thomas:13} Thomas, J., Saglia, R.~P., Bender, R., Erwin, P., \& Fabricius, M.\ 2013, arXiv:1311.3783 


\bibitem[Toomre \& Toomre(1972)]{TT:72} Toomre, A., \& Toomre, J.\ 1972, \apj, 178, 623 

\bibitem[Toomre(1964)]{Toomre:64} Toomre, A.\ 1964, \apj, 139, 1217 


\bibitem[Tremaine et al.(1975)]{TOS:75} Tremaine, S.~D., Ostriker, J.~P., \& Spitzer, L., Jr.\ 1975, \apj, 196, 407 

\bibitem[Trujillo et al.(2004)]{Trujillo:04} Trujillo, I., Erwin, P., Asensio Ramos, A., \& Graham, A.~W.\ 2004, \aj, 127, 1917 

\bibitem[Turner et al.(2012)]{Turner:12} Turner, M.~L., C{\^o}t{\'e}, P., Ferrarese, L., et al.\ 2012, \apjs, 203, 5 



\bibitem[Walcher et al.(2005)]{Walcher:05} Walcher, C.~J., van der Marel, R.~P., McLaughlin, D., et al.\ 2005, \apj, 618, 237 


\end{thebibliography}
\end{document}